\documentstyle[11pt,newpasp,twoside,epsf]{article}
\def\edcomment#1{\iffalse\marginpar{\raggedright\sl#1\/}\else\relax\fi}
\reversemarginpar

\begin{document}
\title{Two formation paths for cluster dwarf galaxies?}
% \author{Bianca M. Poggianti}
%\affil{Osservatorio Astronomico di Padova - Italy}
\author{Bianca M. Poggianti, Nobunari Kashikawa, Terry Bridges, Bahram Mobasher, Yutaka Komiyama, Dave Carter, Sadanori Okamura, Masafumi Yagi}
%\author{Terry Bridges}
%\author{Bahram Mobasher}
%\author{Yutaka Komiyama}
%\author{Dave Carter}
%\author{sadanori Okamura}
%\author{Masafumi Yagi}
%\affil{Their affiliation}

\begin{abstract}
A surprising result of our recent spectroscopic survey of galaxies in the Coma cluster has been the discovery of a possible bimodal distribution in the metallicities of faint galaxies at $M_B>-17$. We identified a group of dwarfs with luminosity-weighted metallicities around solar and a group with [M/H] around -1.5. A metallicity bimodality among galaxies of similar luminosities is unexpected and suggests that faint cluster galaxies could be an heterogeneous population that formed through more than one evolutionary path, possibly as a consequence of the cluster environment.
\end{abstract}

\section{The origin of dwarf galaxies in clusters}
A correlation between metallicity and mass (or luminosity) of galaxies has been established in numerous galaxy samples. including luminous early-type galaxies (e.g. Kobayashi \& Arimoto 1999, Kuntschner 2000), spirals and irregular galaxies (Zaritsky et al. 1994, Lee et al. 2003), cluster dwarfs (Brodie \& Huchra 1991) and the Local group (Mateo 1998).
The fact that more luminous galaxies are more metal-rich is believed to be the main driver of a number of observed correlations, including the color-magnitude relation followed by early-type galaxies in clusters (Kodama et al. 1998).

%\section{The early results}
In the last years we have carried out a photometric and spectroscopic study of galaxies in the Coma cluster, obtaining William Herschel Telescope spectra for about 300 cluster members including both giant and dwarf galaxies down to $M_B \sim -14$.
One of the results of the spectral analysis was the discovery of a bimodal distribution in the metallicities of the faint cluster galaxies with signs of recent star formation (Poggianti et al. 2001). We identified a group of dwarfs with metallicity around solar or even supersolar (hereafter, metal-rich) and a group with [M/H] around -1.5 (metal-poor). All of these dwarfs have luminosity weighted ages of 3 Gyr or less. The metallicity and age estimates were based on the analysis of metallicity-sensitive and age-sensitive spectral indices in the Lick system.

Given the faintness of these galaxies and the relatively low signal-to-noise of our William Herschel spectra, errorbars on spectral indices and on the age and metallicity estimates were too large to draw any definite conclusion. Two additional results, however, lead us to believe that the metallicity bimodality could be real. First of all, in Poggianti et al. (2001) we obtained a coadded spectrum of the metal-rich and the metal-poor dwarfs separately and we showed that, in addition to the magnesium indices on which the metal-rich versus metal-poor division was based, most of the other metallicity-sensitive indices were indeed stronger in the average spectrum of the metal-poor galaxies.
Secondly, the two groups of galaxies also clearly separate in the color-magnitude diagram, with the metal-poor dwarfs being significantly bluer than the metal-rich ones {\it at similar luminosities}.

%\section{The Subaru spectra}
In order to either confirm or disprove the existence of a metallicity spread, we have started a multislit programme with FOCAS at the Subaru telescope to obtain high signal-to-noise spectra.
Only a few spectra of sufficient quality could be obtained so far due to weather conditions during the first Subaru run. 

Line indices have been measured and are currently being compared with spectrophotometric models. These first measurements are encouraging, confirming that metallicity-sensitive indices are indeed very different, despite the similar magnitudes of these galaxies. However, only the simultaneous analysis of several line indices will establish whether this corresponds to large variations in metal content or/and ages or/and chemical abundance ratios. 

%\section{Implications}
At this point the existence of a bimodal distribution of metallicities among faint galaxies in Coma is an intriguing possibility suggested by spectral indices and supported by colors. 
Should the metallicity spread at faint magnitudes be confirmed by the high signal-to-noise spectra, it would have far reaching consequences for our understanding of early-type galaxy formation and evolution, and the hypothesis of different formation processes for faint galaxies in clusters should be seriously contemplated.

The metal-poor dwarfs could represent a ``truly primordial'' population, while the metal-rich ones could be the evolved remnants of more massive galaxies that lost some of their mass (and luminosity) due to environmental effects. Alternatively, the metal-rich dwarfs could be in some way related with another class of dwarf galaxies that are known to deviate from the metallicity/luminosity relation, created from the tidal debris of mergers between metal-rich massive galaxies (tidal dwarfs). These and other speculative options are currently open.

%If the origin of one of te two types of dwarf populations were connected to environmental effects transforming more massive galaxies, then the deficit of dwarf galaxies in clusters compared to the numbers predicted by hierarchical models of galaxy formation would be even more severe than currently believed.

%\begin{figure}[h]
%\plottwo{poggiantifig1.ps}{poggiantifig2.ps}
%\caption{{\bf Left -} Color-magnitude diagram of metal-rich and metal-poor young dwarfs (top panel). For comparison, the color-magnitude diagram of all Coma galaxies is shown in the bottom panel. {\bf Right -} Subaru spectra of two metal-rich and two metal-poor dwarfs, dereshifted at z=0 and binned at 6 \AA.}
%\end{figure}

%\acknowledgments{I thank everyone ....}

\end{document}